# A NEW TECHNIQUE FOR MEASURING THE STRANGENESS CONTENT OF THE PROTON ON THE LATTICE

Jeffrey E. Mandula

*Department of Energy, Division of High Energy Physics, Washington, DC 20585*

and

Michael C. Ogilvie

*Department of Physics, Washington University, St. Louis, MO 63130*

**ABSTRACT**

A new technique for computing the strangeness content of the proton on the lattice is described. It is applied to the calculation of the strange quark contribution to the proton's spin, specifically to the evaluation of the matrix element $\langle ps | \bar{s}\gamma_\mu\gamma_5 s | ps \rangle$. Preliminary results are not in disagreement with the EMC experiment.

The EMC experiment measuring the $g_1$ structure function in polarized deep inelastic muon proton scattering gave the surprising result that the strange quark contribution to the proton spin was substantial[1]. Their measurement,

$$\int_0^1 g_1(x)\,dx = .126 \pm .010 \pm .015 \tag{1}$$

when combined with information from neutron $\beta$ decay and hyperon semileptonic decays gives, for the strange quark content of the proton's spin,

$$\Delta s = -.19 \pm .07 \tag{2}$$

The error includes both the experimental errors and an estimate of the theoretical uncertainty coming from the use of flavor SU(3) symmetry in the analysis. In view of the fact that the proton has no strange valence quarks, this is a surprisingly large value when compared to the nonstrange quark content inferred from the same measurements

$$\begin{aligned}\Delta u &= +.74 \pm .05 \\ \Delta d &= -.51 \pm .05\end{aligned} \tag{3}$$

The proton sigma term also has an anomalously large strange quark contribution. An analysis of $\pi N$ scattering leads to the conclusion[2]

$$\frac{\langle p | \bar{s}s | p \rangle}{\langle p | \bar{u}u + \bar{d}d + \bar{s}s | p \rangle} \sim 0.2 \tag{4}$$

Both of these results suggest that strange quark operators may have fairly large matrix elements between proton states. This violates expectations based on the naive quark model and the Zweig rule. This suggests that there are processes for which the valence quark model is inadequate, and the effects of the "sea" are substantial.

It is clearly desirable to carry out a lattice simulation of the relevant matrix elements. Lattice calculations are capable of providing first principles theoretical results based on QCD alone. If successful, this would provide one of the rare cases in which QCD could explain experimental results beyond the ken of the naive quark model.

The matrix element which expresses the strange quark contribution to the proton's spin,

$$2m s_\mu \Delta s = \langle ps | \bar{q}_s i \gamma_\mu \gamma_5 q_s | ps \rangle \tag{5}$$

can be extracted from a lattice evaluation of the three point function

$$\Delta s = \lim_{\substack{x_4 \to -\infty \\ z_4 \to +\infty \\ (y_4 = 0)}} \frac{\sum_{\vec{z},\vec{y}} Tr\, i\gamma_i\gamma_5 P_+ \langle 0 | \Psi(z) \bar{s}(y) i\gamma_i\gamma_5 s(y) \bar{\Psi}(x_4,0) | 0 \rangle}{\sum_{\vec{z}} Tr \langle 0 | \Psi(z) \bar{\Psi}(x_4,0) | 0 \rangle} \quad (6)$$

where $\Psi(x)$ is a proton field operator constructed from nonstrange quark fields and $P_+$ is the projection operator on positive energy states. The quark line flow of the three point function is illustrated in Figure 1.

The major computational difficulty in calculating this three point function lies in the fact that the strange quark loop must be evaluated at each point on a time slice in order that the momentum transfer to the $\bar{q}_s i\gamma_\mu\gamma_5 q_s$ operator may be constrained to be zero. Each such evaluation involves solving the Dirac equation on the lattice, well known to be a very computationally intensive activity. This is in addition to the computer time necessary to create the lattice field configurations and compute the quark propagators needed to construct the proton.

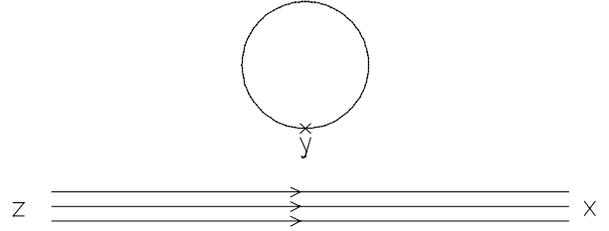

**Figure 1** Calculation of the strange quark loop correlation with the proton propagator

Two observations may make such calculations feasible. One is that it not really necessary to compute the closed loop on every site on a time slice. For example, if one were to do such computations on alternate sites in each direction (saving a factor of 8 in computation), the effect would be to include an admixture of the largest momentum in each direction, $\pi/a$, along with the zero momentum component. All of such high momentum components are suppressed by the exponential decay of the proton propagator. At large Euclidean time, the three point function falls off proportional to $exp(-E(p)t)$, where $E$ is the lattice energy of he proton state with momentum $p$, and $t$ is the interval between the proton source point and the $\bar{q}_s i\gamma_\mu\gamma_5 q_s$ operator.

The second observation is that the solution to a lattice propagator equation at zero separation can be rapidly convergent. Consider the one dimensional Klein-Gordon equation for the propagator on a lattice of $2N + 1$ points satisfying Neumann boundary conditions:

$$\phi(n+1) - 2\phi(n) + \phi(n-1) - m^2\phi(n) = \delta(n)$$

$$\phi(N) = \phi(N-1)$$
$$\phi(-N) = \phi(-N+1)$$

(7)

Its exact solution at the origin is

$$\phi(0) = \frac{1 + e^{(1-2N)M}}{2e^{-M} - (2+m^2) + e^{(1-2N)M}(2e^M - (2+m^2))} \tag{8}$$

where $\sinh(M/2) = m/2$. This converges exponentially in $N$, the extent of the lattice, to its $N \to \infty$ value,

$$\phi_{[N \to \infty]}(0) = \frac{1}{2e^{-M} - (2+m^2)} \tag{9}$$

This is a general property of the irrelevance of boundary conditions in the large volume limit. It can be understood intuitively from the perspective of a hopping parameter expansion, in which $2N$ steps are required to hop from the origin to the boundary and back, giving finite $N$ corrections proportional to $\kappa^{2N}$.

Each of these observations leads to an approximation which is capable of saving orders of magnitude in computation. Furthermore, both of them can be systematically improved. Adding additional points to the measurement of the operator decreases the contamination of the zero momentum component with high momentum contributions. Increasing the size of the box on which the propagator equation is solved improves the accuracy of the result exponentially.

Let us apply these ideas to the evaluation of the strange quark fraction of the proton spin. To evaluate the Green's functions, we have used a set of 16 quenched lattice field configurations of size $16^3 \times 24$ at $\beta = 5.7$, provided at NERSC by the Bernard-Soni collaboration.[3] Quark propagators using Wilson fermions are available with these lattices, so that the principal computational task is the calculation of the strange quark loop. The strange quark operator is evaluated on a time slice 5 units from a pointlike proton source, in an attempt to balance the need for an isolated proton state against the requirements of low noise.

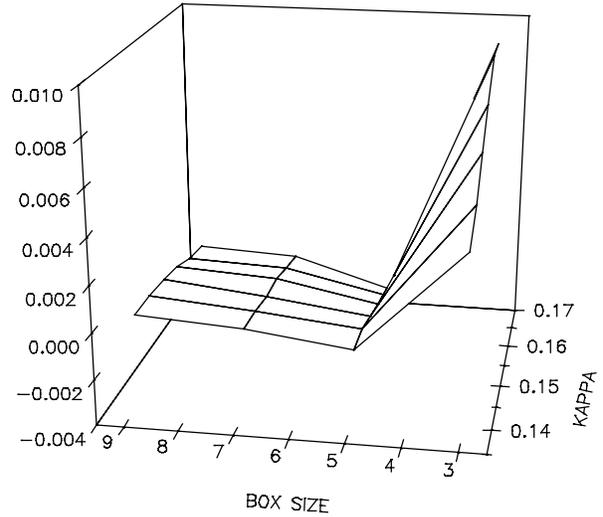

**Figure 2** Convergence of the strange quark loop at a single site as a function of box size $N$ for several values of the hopping constant $\kappa$

An enormous savings in computational time is realized if the strange quark loop is evaluated in a box substantially smaller than the entire lattice. However, the value so obtained should be an accurate estimate of the value which would be obtained were the entire lattice used.

We have explored this issue for six values of κ using boxes of size 3,5,7 and 9. The conjugate gradient method was used to perform the inversion. The results are shown below in Figure 2. For a range of κ values from 0.14 to 0.165, the value of the operator $\bar{q}_s i\gamma_\mu\gamma_5 q_s$ shows good convergence to a stable value by the time the box size reaches 9. We have subsequently used a box with $9^4$ sites for each conjugate gradient inversion. The kappa value used for the strange quark was κ = .162.

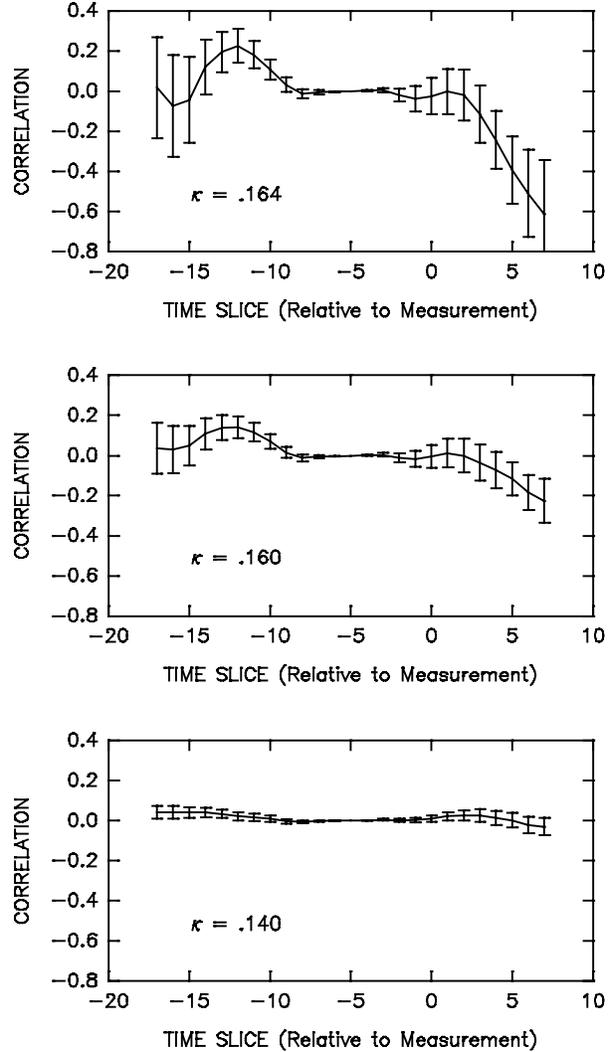

Three values of the hopping constant were used for the valence quarks, κ = .140, .160, and .164. On these lattices, the critical value of the hopping constant is $\kappa_c \approx .170$. Using a linearized relation between κ and the bare quark mass $2ma = 1/\kappa - 1/\kappa_c$, these correspond to quark masses of about .1, .2, and .6 GeV, respectively. The value of the hopping constant used for the strange quark in the loop corresponds to a mass of about .15 GeV.[4]

For each of the 16 gauge field configurations, the strange quark loop was evaluated on a fixed time slice for every other site in each direction. This results in a mixing of the desired zero momentum component of the operator with components of momentum zero or π/a in each direction. The magnitude of this contamination for each of the seven high momentum states is $\exp(-(E(p)-E(0))t)$. A simple free lattice propagator estimate of this gives $\exp(-(\sqrt{M^2+4d}-M)t)$ where d is 1,2, or 3, depending on the number of directions along which the momentum is nonzero. For the present calculation, where $M \approx 1.5$ and $t = 5$, This gives suppressions of at least $e^{-5} \approx .01$.

**Figure 3** The strange quark spin fraction of the proton Δs, for three values of the proton valence quark hopping constant, κ = .164, .160, and .140

Figure 3 shows the results of the calculation for three values of the valence quark hopping constant κ. The x axis gives the location of the time slice on which the proton is detected, relative to the time slice on which the strange quark operator is evaluated.

Ideally, the signal would appear when the proton detection slice moves past the strange quark operator, and it would rise asymptotically to a stable value.

From these results it is not possible to infer a quantitative value of the strange quark's contribution to the proton spin. However, they do give a strong hint that the strange quark spin fraction is negative and substantial. At the lightest value of the valence quark mass, corresponding to $\kappa = .164$, $\Delta s$ systematically departs from zero as the time at which the final proton is detected grows, becoming 2 standard deviations below zero at the largest values of $t$. However, no plateau in $\Delta s$ has been reached. That the departure form zero is not purely a statistical fluctuation is indicated by the contrast with the negative $t$ side of the $\kappa = .164$ graph in Figure 3, where no signal is theoretically expected, and no hint of one appears.

The graph also shows what would be required to carry out a quantitative evaluation of $\Delta s$. The problem is that below $t = 5$ units from the proton source, which is about as far as one can go before the statistical errors become prohibitively large, the contamination of the proton propagator with higher mass states is considerable. One can get a quantitative estimate of the magnitude of the admixture of high mass contributions by directly examining the proton propagator. Figure 4 shows the proton propagator, together with its single proton component estimated by fitting to the tail of the curve. At $t = 4$, about half of the propagator's value is due to the contributions of high mass contributions.

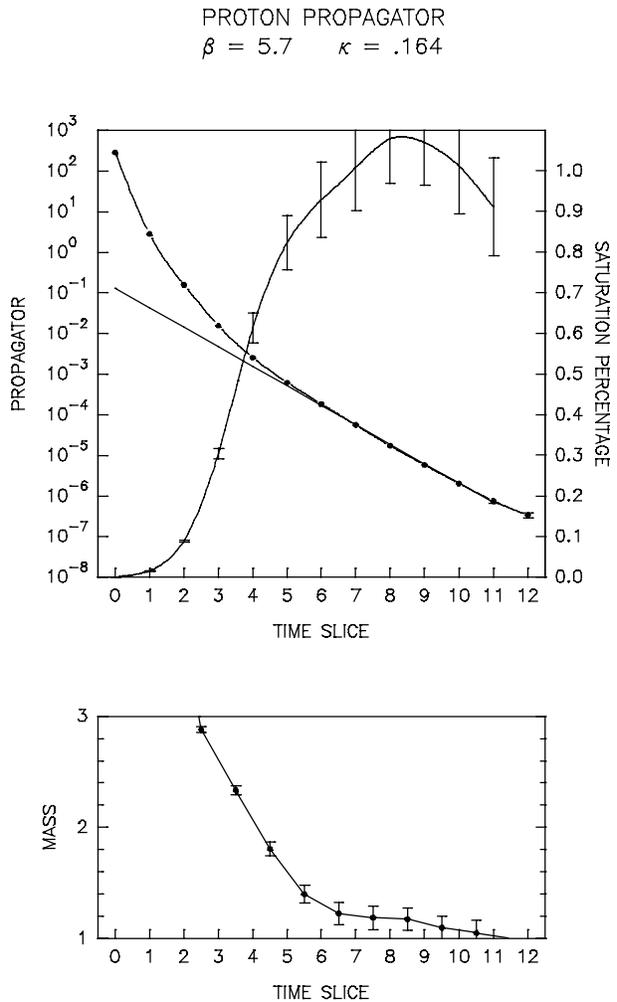

**Figure 4** The proton propagator and its percentage saturation by the single proton contribution; the effective mass between each pair of time slices

In order to extract quantitative information from this sort of analysis, one would need either proton propagators determined over a larger range of time separations, or propagators constructed from operators which project on the single proton state much more efficiently. In either case, what would be required would be for the value of the correlation determining $\Delta s$ to be stable over several time slices.

As expected from the naive quark model, there is no indication of any strangeness content in a proton whose valence quarks are taken as very heavy. Only for the lightest u and d valence quarks in the proton is there some indication of a signal. The appearance and magnitude of the signal in the present non-optimized calculation gives reason to believe that the strange quark spin content of the proton can be accurately calculated using lattice techniques.

This calculation could be improved upon in a number of ways. As is usual in lattice calculations, the investment of computational resources allows for many improvements, *e.g.*, larger volumes and smaller lattice spacing. It is well known that lattices of this size and value of $\beta$ are suspect for the study of weak interaction matrix elements.[5] A better calculation of $\Delta s$ would be performed most sensibly as a part of a larger investigation in which other hadronic properties are studied. Obviously, it would be interesting to compare results from a simulation of full QCD with results from the quenched approximation. There are additional improvements which carry little cost in computation. Improved proton sources isolate the proton after fewer time steps than pointlike sources; the use of such sources is now standard. As emphasized by El-Khadra *et al.*, some hadronic quantities, such as the $J/\psi$-$\eta_c$ splitting, are sensitive to $O(a)$ terms in the Wilson action.[6] It is very likely that the spin content of the proton is also sensitive to these terms. This problem can be addressed using the improved cloverleaf action.[7] Finally, we note that a finite renormalization of the matrix element is necessary to explicitly compare the lattice result with continuum measurements. We have not included this factor here because our results do not constitute a quantitative evaluation of $\Delta s$, but it will be necessary in later lattice calculations which are to be compared with experiment.

**Acknowledgements**

The authors would like to thank Claude Bernard and Amargit Soni for generously making publically available the lattices used in this analysis. This work has been supported in part by the U.S. Department of Energy under Grant No. DE-FG02-92ER40626.

**References**

1. The EMC Collaboration, Nucl. Phys. **B328**, 1 (1989).

2. T.P. Cheng, Phys. Rev. D **13**, 2161 (1976); J.F. Donoghue and C. Nappi, Phys. Letts. **168B**, 105 (1986); B.L. Ioffe and M. Karliner, Phys. Letts. **247B**, 387 (1990).

3. See, for example, C. Bernard, T. Draper, R. Hockney, A. Rushton, and A. Soni, Phys. Rev. Letts. **55**, 2770 (1985).

4. C. Bernard, A. El-Khadra, and A. Soni, Phys. Rev. D **43**, 2140 (1991).

5. C. Bernard, Proceedings of the TASI '89 Summer School (*Weak Matrix Elements On and Off the Lattice*).

6. A. El-Khadra, G. Hockney, A. Kronfeld and P. Mackenzie, FERMILAB-PUB-91/354-T; P. Mackenzie, FERMILAB-PUB-92/09-T; A. El-Khadra, FERMILAB-PUB-92/10-T.

7. J.E. Mandula, J. Govaerts, and G. Zweig, Nucl. Phys. **B228**, 109 (1983); B. Sheikholeslami and R. Wohlert, Nucl. Phys. **B259**, 572 (1985).